\documentclass[amsmath,amssymb,10pt,letterpaper,nofootinbib,balancelastpage,notitlepage,superscriptaddress,floatfix,preprintnumbers]{revtex4-2}

\usepackage{graphicx}	
\usepackage{ragged2e}	
\usepackage{bm}		    
\usepackage[sort&compress]{natbib}	 	
	\setcitestyle{square,numbers,comma}	
\usepackage[colorlinks=true,urlcolor=blue,linkcolor=black,citecolor=red]{hyperref}	
\usepackage{slashed}
\renewcommand{\eqref}[2][]{Eq{#1}.~(\ref{eq:#2})}		
\newcommand{\orcid}[1]{\href{https://orcid.org/#1}{\,\includegraphics[width=8px]{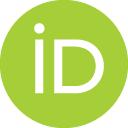}}}

\def\gae{g_{ae}}

\usepackage{color}
   \definecolor{k}{RGB}{0,0,0}
   \definecolor{r}{RGB}{255,0,0}
   \definecolor{R}{RGB}{255,0,0}
   \definecolor{o}{RGB}{255,94,1}
   \definecolor{O}{RGB}{255,165,0}
   \definecolor{p}{RGB}{148,55,255}
   \definecolor{P}{RGB}{148,55,255}
   \definecolor{g}{RGB}{0,150,50}
   \definecolor{G}{RGB}{0,150,50}
   \definecolor{b}{RGB}{0,0,255}
   \definecolor{B}{RGB}{0,0,255}
\usepackage{ezedits}

\begin{document}

\title{Electron $g-2$ corrections from axion dark matter}
\date{\today}
\author{Ariel Arza\orcid{0000-0002-2254-7408}}
\email{ariel.arza@gmail.com}
\affiliation{Department of Physics and Institute of Theoretical Physics, Nanjing Normal University, Nanjing, 210023, China}
\affiliation{Tsung-Dao Lee Institute, Shanghai Jiao Tong University, Shanghai 200240, China}
\author{Jason L. Evans}
\email{jlevans@sjtu.edu.cn}
\affiliation{Tsung-Dao Lee Institute, Shanghai Jiao Tong University, Shanghai 200240, China}

\begin{abstract}
We consider the effects of a local axion dark matter background on $g-2$ of the electron. We calculate loop corrections to the photon-electron vertex and determine analytical formulas for the spin and cyclotron frequencies when the electron is in an external magnetic field. By comparing with current measurements of these observables, we are able to place the strongest constraint on the axion-elecron coupling for axion masses below $3\times10^{-18}\text{eV}$. 
\end{abstract}

\maketitle

\section{Introduction} \label{intro}
The identity of dark matter is currently one of the most debated puzzles in all of natural science. The QCD axion \cite{Peccei:1977hh,Kim:1979if,Shifman:1979if,Dine:1981rt,Zhitnitsky:1980tq} and axion-like particles (ALPs) \cite{Svrcek:2006yi,Arvanitaki:2009fg} are among the leading candidates \cite{Preskill:1982cy,Abbott:1982af,Dine:1982ah,Arias:2012az}. Axions are commonly searched for by looking at their interaction with photons, especially axion to photon conversions in strong external magnetic fields \cite{Sikivie:1983ip}. See Refs. \cite{Sikivie:2020zpn} and \cite{Irastorza:2018dyq} for recent reviews on axion phenomenology.

For axion-electron interactions, strong constraints are placed on masses between $0.1$ and $10\,\text{keV}$ by the underground electron recoil experiments such as XENON1T \cite{XENON:2019gfn} and XENONnT \cite{XENON:2022ltv}, while for masses above $10\, \text{keV}$, strong constraints are from radiative decays of axion dark matter via a one loop electron diagram \cite{Ferreira:2022egk}. For masses below $0.1\,\text{keV}$, the constraints on the axion-electron coupling $g_{ae}$ are dominated by the resulting energy loss of red-giant branch stars, which has been updated to $g_{ae}<1.3\times10^{-13}$ \cite{Capozzi:2020cbu}.

In this work we constrain the axion dark matter interaction with the electron by computing the effects of a cold axion background on the electron magnetic dipole moment. As it was pointed out recently in \cite{Evans:2023uxh}, ultra light bosonic dark matter corrects the value of $g-2$ of the electron substantially due to the high occupancy number of the background field. It was shown that the corrections become larger for lower masses, allowing for a constraint on the dark photon kinetic mixing parameter of $\chi<7.1\times10^{-11}\left(m_{\gamma'}\over10^{-14}\,\text{eV}\right)$. This being the strongest constraint on dark photon dark matter for dark photon masses $m_{\gamma'}$ below $10^{-14}\,\text{eV}$. 

In the same spirit, we expect constraints on the axion-electron coupling. We could estimate a constraint to the axion-electron coupling from the naive mapping $\chi\leftrightarrow g_{ae}$. However, as scalar have zero spin, the $g-2$ corrections should be suppressed by the axion velocity. As shown in our results, we get the constraint to be $g_{ae}\lesssim4.48\times10^{4}\left(m_a/\text{eV}\right)$, which is competitive with current constraints for masses close to $10^{-17}\,\text{eV}$, and becomes even stronger for lower masses. We also take this opportunity to include in our analysis other dark matter candidates with a standard Yukawa type interaction.

This article is organized as follows: in Sec. \ref{formalism} we discuss the formalism for the loop correction inspired by QFT techniques at finite temperature and give a fairly general discussion on the corrections to the electron-photon vertex in the presence of a background. In Sec. \ref{results}, we apply this formalism for the case of an axion dark matter background. We consider axions with both a derivative term interaction and also a pseudoscalar Yukawa interaction with the electron\footnote{Although in the literature both interaction are considered as equivalent, this is only guaranteed to be true for on-shell vertices.}. We also consider a dark scalar interacting with a Yukawa coupling. In Sec. \ref{pheno} we compute the cyclotron and spin frequency of an electron in a uniform magnetic field, and compare our results with an updated measurement of these observables in a penning trap. We finally conclude in Sec. \ref{conclusion}.

\section{Loop corrections formalism} \label{formalism}

The zeroth order contribution to the electron magnetic moment arises from the fermion-photon vertex in quantum electrodynamics (QED). The one loop corrections to the classical QED theory provides a contribution given by $g-2=\alpha/\pi$, where $\alpha$ is the fine structure constant. For this case, the loops only involve virtual photons.

One can, of course, include corrections from beyond SM particles that interact with the electron. In this work we consider three different interactions, an axion with a derivative coupling to fermions, a pseudoscalar with a Yukawa type interaction (pYukawa) and a scalar with a Yukawa coupling (sYukawa), and calculate the effects of each interaction separately. Thus, our interaction Lagrangian in momentum space has the form
\begin{align}
{\cal L}_I=\gae^2 a(q) \bar u(k)\hat\Gamma(q) u(k) \label{eq:lag}
\end{align}
where $a(q)$ is the axion, pseudoscalar or scalar field with momentum $q$, $u(k)$ the electron spinor with momentum $k$, and $\hat\Gamma(q)$ a model dependent matrix that accounts for the type of interaction and that is written as 
\begin{align}
\hat\Gamma(q)=
\begin{cases}
-{i\slashed{q}\over2 m_e}\,\gamma_5 & \text{derivative}
\\
-\,i\gamma_5 & \text{pYukawa}
\\
~~~\,1         & \text{sYukawa}
\end{cases}, \label{eq:models}
\end{align}
where $\slashed{q}=q^\mu\gamma_\mu$, and $\gamma_\mu$ and $\gamma_5$ are Dirac gamma matrices. The coupling constant $\gae$ can be written with the same symbol for the three interactions since they are considered independently.

For virtual particles of the above models, the contributions to the anomalous magnetic moment are of the order of $g-2\sim{\gae^2\over16\pi^2}$. This is too small even for comparison with the highest precision measurements. Indeed, the values of $\gae$ that gives a $g-2$ contribution comparable with current uncertainties, are already constrained by many other experiments and observations (See Sec. \ref{intro}).

However, if these scalar or pseudoscalar particles are not virtual particles but are instead real non-relativistic particles of a background, things are different. Based on the work \cite{Evans:2023uxh}, we anticipate that, for the case of a local dark matter background, the $g-2$ correction picks up a factor of order\footnote{This estimate comes from the occupation number, which roughly scales as $\rho_\mathrm{dm}\over m_{dm}^4$, multiplied by the expected IR cut off, which is of order $m_\mathrm{dm}^2/m_e^2$.} $\sim{\rho_\mathrm{dm}\over m_\mathrm{dm}^2m_e^2}$ with respect to the case with virtual particles. Here $m_e$ is the electron mass, $m_\mathrm{dm}$ the dark matter candidate mass, and $\rho_\mathrm{dm}$ is the local energy density (in \cite{Evans:2023uxh}, the last two parameters would correspond to those of a dark photon dark matter candidate). Assuming that these particles make up all the dark matter energy density, i.e. $\rho_\mathrm{dm}=0.3\,\text{GeV}/\text{cm}^3$ for the standard halo model, this additional factor results in an enhancement, with respect to the virtual particle case, if $m_\mathrm{dm}$ is below $\sim10^{-9}\,\text{eV}$.

\begin{figure}[t]
  \includegraphics[width=0.4\linewidth]{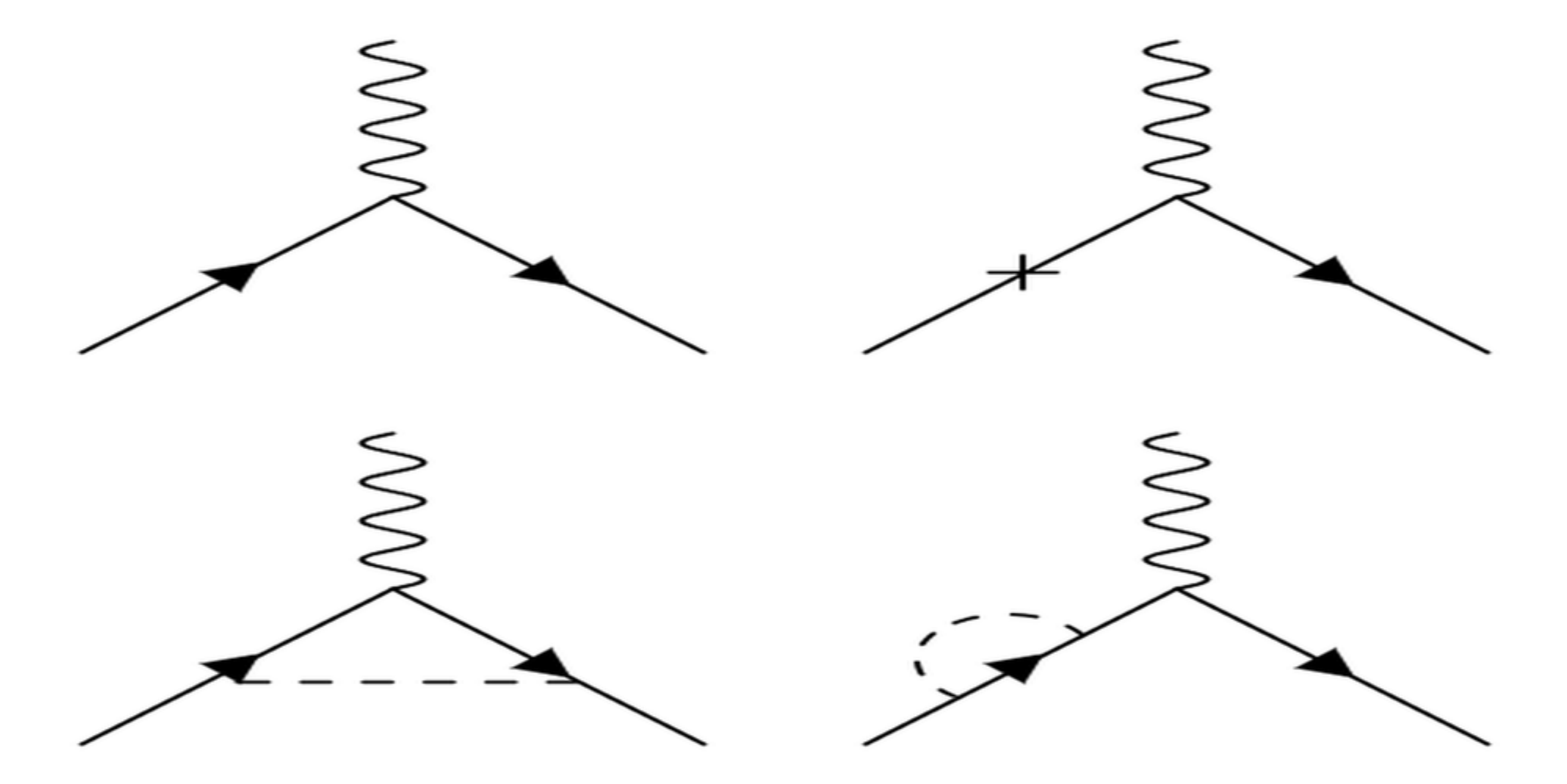}
\caption{Relevant diagrams for our computations. The dashed lines correspond to the dark matter background states.} \label{fig:diagrams}
\end{figure}

We consider the case in which only one of the candidates from Eq. (\ref{eq:lag}) and Eq. (\ref{eq:models}) makes up 100\% of the dark matter, then we consider each case separately and denote its mass and energy density by $m_a$ and $\rho_a$, respectively.
For calculating the dark matter background effects we base our analysis on the techniques discussed in Refs. \cite{Donoghue:1984zz,Donoghue:1983qx} for QED at finite temperature and \cite{Evans:2023uxh} for a dark matter background. In a background, the propagator of a scalar or pseudoscalar dark matter field is corrected in the same way it is in finite temperature,
\begin{equation}
\Delta_F(q)={1\over q^2-m_a^2+i\epsilon}-2\pi if_a(q)\delta(q^2-m_a^2)
\end{equation}
except $f_a(q)$ is the occupancy number of the background dark matter field, and not one of the standard thermal equilibrium distributions. 

The electron self-energy for a particular momentum $k^\mu\equiv(E,\vec k)$ is now $\Sigma(k)=\Sigma_0(k)+\Sigma_\beta(k)$, where $\Sigma_0(k)$ corresponds to the electron self-energy in the vacuum and $\Sigma_\beta(k)$ the background contribution given by
\begin{align}
\Sigma_\beta(k) &= \int{d^4q\over(2\pi)^3}f_a(q)\delta(q^2-m_a^2)i\hat\Gamma(q){1\over\slashed k+\slashed q-m_e}i\hat\Gamma(q)
\\
&=\int{d^4q\over(2\pi)^3}f_a(q)\delta(q^2-m_a^2){i\hat\Gamma(q)(\slashed k+\slashed q+m_e)i\hat\Gamma(q)\over(k+q)^2-m_e^2}~~.\label{eq:selfen0}
\end{align}
It can also be expressed in general terms as
\begin{equation}
\Sigma_\beta(k)=B(k)+C(k)(\slashed{k}-m_e)+\slashed{D}(k) \label{eq:selfen1}
\end{equation}
where $B$, $C$ and $D_\mu$ are functions determined by the type of interactions we are considering. They are written in detailed in Sec. \ref{results}.  The form of this correction to the electron self-energy highlights the fact that the background gives a preferred frame and so violates Lorentz invariance. Without this violation, $\slashed{D}(k)$ would be proportional to $\slashed{k}$ and would contribute to the standard wave function and mass renormalization. However, because of this term, the background corrects the spinor dynamics in a non-trivial way, 
\begin{equation}
(\slashed{k}-m_e+\Sigma_\beta(k))u_\beta(k,s)=0, \label{eq:spinor1}
\end{equation}
with spinors which are normalized in the usual way $u_\beta(k,s)^\dagger u_\beta(k,s')=\delta_{ss'}$. We define $\tilde k^\mu$ and $\tilde m_e$ as
\begin{align}
\tilde{k}^\mu & = (1+C(k))k^\mu+D^\mu(k) \label{eq:defktilde}
\\
\tilde m_e & = (1+C(k))m_e-B(k)     \label{eq:defmtilde}
\end{align}
such that the spinor dynamical equation is written as
\begin{equation}
(\slashed{\tilde{k}}-\tilde m_e)u_\beta(k,s)=0. \label{eq:spinor2}
\end{equation}
From this, the spins sums are now
\begin{align}
\sum_su_\beta(k,s)\bar u_\beta(k,s)&={\tilde{\slashed{k}}+\tilde m_e\over\tilde E} \label{eq:spinsum1}
\\
\sum_sv_\beta(k,s)\bar v_\beta(k,s)&={\tilde{\slashed{k}}-\tilde m_e\over\tilde E}. \label{eq:spinsum2}
\end{align}
The electron energy $E$ is determined by finding the values of $k^0$ that satisfy $\tilde k^2-\tilde m_e^2=0$. We get (See appendix \ref{app:renorm}) $k^0=\pm E$, where
\begin{equation}
E(k)=\sqrt{\omega_k^2-2\,k\cdot D(k)-2\,m_e\,B(k)}~~, \label{eq:E1}
\end{equation}
where we have defined
\begin{equation}
\omega_k=\sqrt{\vec k^2+m_e^2}~~. \label{eq:omega1}
\end{equation}

As is clear from this discussion, the wave function is of course renormalized by the background as well.  The renormalized Feynman propagator takes on the usual form 
\begin{equation}
S_F^R(x-x')=\int {d^4k\over(2\pi)^4}e^{-ik\cdot(x-x')}{Z_2^{-1}\over\tilde{\slashed{k}}-\tilde m_e+i\epsilon} \label{eq:elecprop1}
\end{equation}
where $Z_2$ is the wave function renormalization constant. On the other hand, the renormalized propagator can also be found as follows
\begin{align}
S_F^R(x-x') &= -i\left<0\right|T[\psi_e(x)\bar\psi_e(x')]\left|0\right> \nonumber
\\
&= -i\int{d^3k\over(2\pi)^3}\left(\Theta(t-t'){\tilde{\slashed{k}}+\tilde m_e\over2\tilde E}e^{-ik\cdot(x-x')}\right. \nonumber
\\
& \left.-\Theta(t'-t){\tilde{\slashed{k}}-\tilde m_e\over2\tilde E}e^{ik\cdot(x-x')}\right)~~. \label{eq:elecprop2}
\end{align}
Comparison of Eq. (\ref{eq:elecprop1}) with Eq. (\ref{eq:elecprop2}) gives the following definition of the wave function renormalization (see appendix \ref{app:renorm} for a detailed derivation)
\begin{align}
Z_2(k)^{-1}=1+C(k)+{m_e\over E}{d\over dE}\left(B(k)+{k\cdot D(k)\over m_e}\right)-{D_0(k)\over E}~~.
\end{align}

Now we can compute all the contributions to the electron anomalous magnetic moment. The relevant diagrams are specified in Fig. \ref{fig:diagrams}. Importantly, the counterterm diagrams must be calculated when considering the background-dependent piece due to the Lorentz violation of the background. The one loop correction to quantum electrodynamics vertex can then be parsed into four pieces: self energy, counterterm, vertex correction, and wavefunction renormalization. Details of diagram computations are found in appendix \ref{app:diagrams}. 

We first consider the wave function renormalization contribution which corresponds to the upper left diagram of Fig. \ref{fig:diagrams}. Each fermion leg picks up a factor $Z_2^{-1/2}$ to give 
\begin{equation}
i{\cal M}_\mu^{WFR}=(-ie)\left(Z_2(k)^{-1}+Z_2(k')^{-1}\over2\right)\bar u(k')\gamma_\mu u(k). \label{eq:ampWFR}
\end{equation}

The self-energy piece is found from calculating the bottom right diagram in Fig. (\ref{fig:diagrams}) and takes the form
\begin{equation}
i{\cal M}_\mu^{SE}=(-ie)\bar u(k')\,\gamma_\mu\left(1-Z_2(k)^{-1}-{1\over\slashed k-m_e}\left(B(k)+\slashed D(k)\right)\right)u(k)+..., \label{eq:ampSE}
\end{equation}
where $Z_2(k)^{-1}$ is evaluated at $k^2=m_e^2$. The loop factor for this contribution takes the same form as that found in Eq. (\ref{eq:selfen1}) and is effectively inserted into this diagram in its generic form. This loop factor is then expanded about $k^2=m_e^2$ keeping only the pieces which will contribute when the electron is on-shell. The ``..." at the end of Eq. (\ref{eq:ampSE}) is the analogous contribution from a diagram with the self-energy loop on the outgoing leg. This can be found from the previous expression by taking $k\to k'$ for all terms within the outer parentheses and also moving the $\gamma_\mu$ to the right. 

The counterterm contribution is readily written down, since it must cancel the pieces of the self energy which diverge when the electron momentum is taken on-shell. It is found to be
\begin{equation}
i{\cal M}_\mu^{CT}=(-ie)\bar u(k')\gamma_\mu{1\over\slashed{k}-m_e}(B(k)+\slashed D(k))\,u(k)+.... \label{eq:ampCT}
\end{equation}
Again, the ``..." indicates the analogous contribution coming from a counterterm on the outgoing leg.

The contribution from these three diagrams together give the following
\begin{equation}
i{\cal M}_\mu^{WFR}+i{\cal M}_\mu^{SE}+i{\cal M}_\mu^{CT}=(-ie)\left(2-{1\over2}\left(Z_2^{-1}(k)+Z_2^{-1}(k')\right)\right)\bar u(k')\gamma_\mu u(k)
\end{equation}

The last piece is the vertex correction diagram. Due to the ward identities, the vertex correction diagram must be related to the derivative of the self energy as follows:
\begin{eqnarray}
i{\cal M}^{VER}_\mu\left|_{k' =k}\right. = -ie\bar u(k)\frac{d\Sigma_\beta}{dk^\mu}u(k)~.
\end{eqnarray}
This allows us to expand the vertex correction diagram in small photon momentum given by $\Delta k=k'-k$.  This expansion is effectively an expansion of the vertex correction in terms of derivatives of the self energy and a piece proportional to the photon momentum. We find
\begin{equation}
i{\cal M}^{VER}_\mu = \nonumber -i\frac{e}{2}\bar u(k')\left(\frac{d\Sigma_\beta}{dk^\mu} + \frac{d\Sigma_\beta}{dk'^\mu}+F_\mu (k,\Delta k)\right)u(k)~,\label{eq:VerCorr}
\end{equation}
where $F_\mu(k,\Delta k)$ cannot be determined from the generic form of $\Sigma_\beta (k)$ and instead depends on the details of the particle dark matter model considered. Its form is given by
\begin{equation}
F_\mu(k,\Delta k)={\gae^2\over2}\int{d^4q\over(2\pi)^3}f_a(q)\delta(q^2-m_a^2){\rho_\mu(k,\Delta k,q)\over((k+q)^2-m_e^2)^2} ~~ , \label{eq:F1}
\end{equation}
where
\begin{align}
\rho_\mu(k,\Delta k,q) &= [i\hat\Gamma(q),\slashed k]\,\gamma_\mu\,\Delta\slashed k\,i\hat\Gamma(q)+i\hat\Gamma(q)\,\Delta\slashed k\,\gamma_\mu\,[i\hat\Gamma(q),\slashed k] \nonumber
\\
& ~~~ +i\hat\Gamma(q)(\slashed q\,\gamma_\mu\,\Delta\slashed k-\Delta\slashed k\,\gamma_\mu\,\slashed q)i\hat\Gamma(q)+2m_ei\hat\Gamma(q)[\gamma_\mu,\Delta\slashed k]i\hat\Gamma(q) ~~ . \label{eq:rho1}
\end{align}


The total vertex is then the sum of these four contributions giving
\begin{equation}
i{\cal M}_\mu^\text{total}(k,k')= (-i e)\bar u(k')\left(\gamma_\mu+Q_\mu(k)+Q_\mu(k')+{i\Delta k_\rho\over2m_e}\sigma^{\nu\rho}{d D_\nu\over dk^\mu}+F_\mu(k,\Delta k)\right)u(k) ~~, \label{eq:Mtot1} 
\end{equation}
where $\sigma_{\mu\nu}={i\over2}[\gamma_\mu,\gamma_\nu]$ and
\begin{equation}
Q_\mu(k)={1\over2}\left({d\over d k^\mu}-\gamma_\mu{m_e\over E}{d\over dE}\right)\left(B(k)+{k\cdot D(k)\over m_e}\right)+\gamma_\mu{D_0(k)\over2E}-{D_\mu(k)\over2m_e} ~~ . \label{eq:Q1}
\end{equation}

Throughout this work, we will assume a spherically symmetric dark matter distribution $f(q)$, which simplifies the calculation. 


To conclude this section we want to shortly discuss the scaling of the expected results. As the electron $g-2$ corrections due to a dark matter background are proportional to the dark matter occupancy number $f_a(q)$, and as the occupancy number for non-relativistic particles scales as $f_a\sim1/m_a^4$, in general one would also expect terms of this type in Eq. (\ref{eq:Mtot1}). However, those terms correspond to the analogue of IR divergences. At the end of the day, these contributions from the different diagrams cancel out and the final results are terms that scale at most as $\sim1/m_a^2$. These cancellations are general properties of finite temperature loop calculations and were also discussed in Refs. \cite{Donoghue:1984zz,Donoghue:1983qx}. Now, our final results that scale as $\sim1/m_a^2$ still show a divergence as $m_a\to0$, however it is a physical consequence of the scaling of the occupancy number, i.e., $f_a\sim1/m_a^4$, which diverge very fast as $m_a\to0$. In refs. \cite{Donoghue:1984zz,Donoghue:1983qx} this physical divergence does not appear because they consider a photon thermal distribution where the occupancy number diverge as $f_\gamma\sim1/|\vec q|$ for $|\vec q|\to0$, not as fast as the cold dark matter occupancy number\footnote{In the case of a photon in a Bose-Einstein distribution, the energy density scales as $T^4$ and so vanishes when $T\to 0$. This leads to the correction to $g-2$ from photons in a Bose-Einstein distribution being finite in the limit $T\to 0$.  Here, we fix the energy density to that of the dark matter energy density for all values of $m_a$. This makes it appear that there is a divergence when $m_a\to 0$.} . Another possible issue is that for very small masses perturbation theory breaks down. However, this is not the case for the parameter space we are considering.

\section{Analytical results for (pseudo)scalar dark matter models} \label{results}


In this section we provide explicit expressions of all the terms in Eq. (\ref{eq:Mtot1}). Comparing Eq. (\ref{eq:selfen0}) with Eq. (\ref{eq:selfen1}), the functions $B(k)$ and $D_\mu(k)$ are given by
\begin{align}
B(k) = & -\gae^2\int{d^4q\over(2\pi)^3}f_a(q)\delta(q^2-m_a^2){m_e(b_0+b_2R)\over(k+q)^2-m_e^2} ~~ , \label{eq:B1}
\\
D_\mu(k) = & -\gae^2\int{d^4q\over(2\pi)^3}f_a(q)\delta(q^2-m_a^2){(d_0+d_2R+\bar d{k\cdot q\over m_e^2})q_\mu\over(k+q)^2-m_e^2} ~~ , \label{eq:D1}
\end{align}
where $R=m_a^2/m_e^2$. The values of the coefficients $\{b_i,d_i\}$ and $\bar d$ depend on whether the dark matter has a derivative, pYukawa or sYukawa interaction with fermions. They are given in Table \ref{tab:coeff} for the models we consider.

\begin{table}[ht]
\caption{} 
\centering 
\begin{tabular}{c c c c c c} 
\hline\hline 
interaction & $b_0$ & $b_2$ & $d_0$ & $d_2$ & $\bar d$ \\ [0.5ex] 
\hline 
derivative & 0 & ${1\over2}$ & 0 & $-{1\over4}$ & $-{1\over2}$ \\ 
pYukawa & 0 & 0 & 1 & 0 & 0 \\
sYukawa & 2 & 0 & 1 & 0 & 0 \\  [1ex] 
\hline 
\end{tabular}
\label{tab:coeff} 
\end{table}
We do not provide the function $C(k)$ since it disappeared after summing all diagram contributions, and therefore, it is irrelevant for our calculations. 

We now expand the term $1/((k+q)^2-m_e^2)$ to first order in $k^2-m_e^2$, keeping only terms up to zero order in $q$ (higher orders are negligible for the masses and momenta considered in this work), we have
\begin{align}
{1\over(k+q)^2-m_e^2} &\simeq {1\over2}{1\over k\cdot q}-{1\over4}{m_e^2R\over(k\cdot q)^2}+{1\over8}{m_e^4R^2\over(k\cdot q)^4}+\left(-{1\over4}{1\over(k\cdot q)^2}+{1\over4}{m_e^2R\over(k\cdot q)^3}-{3\over16}{m_e^4R^2\over(k\cdot q)^4}\right)(k^2-m_e^2) ~~. \label{eq:exp1}
\end{align}
As this expansion will be implemented in Eq. (\ref{eq:B1}) and Eq. (\ref{eq:D1}), as well as in the rest of this paper, it is worth defining the next set of integrals
\begin{align}
J &= {\gae^2\over m_e^2}\int{d^4q\over(2\pi)^3}f_a(q)\delta(q^2-m_a^2) ~~ , \label{eq:J}
\\
I_A(k) &= \gae^2\int{d^4q\over(2\pi)^3}f_a(q)\delta(q^2-m_a^2){m_e^2\over(q\cdot k)^2} ~~ , \label{eq:IA}
\\
\bar I_A(k) &= \gae^2\int{d^4q\over(2\pi)^3}f_a(q)\delta(q^2-m_a^2){m_e^6\over(q\cdot k)^4} ~~ , \label{eq:IAbar}
\\
I_\mu(k) &= {\gae^2\over m_e}\int{d^4q\over(2\pi)^3}f_a(q)\delta(q^2-m_a^2){q_\mu\over q\cdot k} ~~ , \label{eq:Imu}
\\
\bar I_\mu(k) &= \gae^2\int{d^4q\over(2\pi)^3}f_a(q)\delta(q^2-m_a^2){m_e^3q_\mu\over(q\cdot k)^3} ~~ , \label{eq:Imubar}
\\
I_{\mu\nu}(k) &= \gae^2\int{d^4q\over(2\pi)^3}f_a(q)\delta(q^2-m_a^2){q_\mu q_\nu\over(q\cdot k)^2} ~~ . \label{eq:Imunu}
\end{align}
Using the expansion in Eq. (\ref{eq:exp1}), the functions $B(k)$ and $D_\mu(k)$ are written as
\begin{align}
B(k) &= {b_0\over4}m_eR\,I_A(k)+\left({b_0+b_2R\over2}I_A(k)+{3\over8}b_0R^2\bar I_A(k)\right)\left(k^2-m_e^2\over2m_e\right) ~~ ,
\\
D_\mu(k) &= -{d_0\over2}m_e\,I_\mu(k)+{1\over2}\left(\bar d\,I_\mu(k)-d_0R\,\bar I_\mu(k)\right)\left(k^2-m_e^2\over2m_e\right) ~~ .
\end{align}
Thus, the term ${dD_\nu\over dk^\mu}$, which appears in Eq. (\ref{eq:Mtot1}), is
\begin{align}
{dD_\nu\over dk^\mu} &= {d_0\over2}I_{\mu\nu}(k)+{k_\mu\over2}\left(\bar dI_\nu(k)-d_0R\,\bar I_\nu(k)\right) ~~ , \label{eq:dD1}
\end{align}
where we have used the fact that ${dI_\nu\over dk^\mu}=-{I_{\mu\nu}(k)\over m_e}$. On the other hand, the terms $Q_\mu(k)$ in Eq. (\ref{eq:Mtot1}), that are defined in Eq. (\ref{eq:Q1}), have the form
\begin{equation}
Q_\mu(k)=S_\mu(k)\left({k_\mu\over m_e}-\gamma_\mu\right)+T_\mu(k) ~~ , \label{eq:Q2}
\end{equation}
where the functions $S_\mu(k)$ and $T_\mu(k)$ are given by
\begin{align}
S(k) &= {1\over4}\left(\bar d\,J+(b_0+(b_2-d_0)R)I_A(k)+{3\over4}b_0R^2\bar I_A(k)\right)
\\
T_\mu(k) &= {1\over4}\left(d_0I_\mu(k)-b_0R\bar I_\mu(k)\right)-{\gamma_\mu\over4}{m_e\over\omega_k}\left(d_0\,I_0(k)-b_0R\,\bar I_0(k)\right) ~~ .
\end{align}
When calculating $Q_\mu(k)+Q_\mu(k')$ we notice that at first order in $\Delta k$, $k_\mu\,S(k)+k_\mu'\,S(k')\simeq{1\over2}(k_\mu+k_\mu')(S(k)+S(k'))$, then using Gordon identity we have $k_\mu\,S(k)+k_\mu'\,S(k')=m_e\gamma_\mu(S(k)+S(k'))-i\sigma_{\mu\nu}\Delta k^\nu S(k)$. On the other hand, we notice that for the term $T(k)+T(k')$, the expansions at first order in $\Delta k$ give vanishing integrals since their integrands are odd in $q$. Combining these two realizations, we get
\begin{equation}
Q_\mu(k)+Q_\mu(k')=-i{\Delta k^\nu\over m_e}\sigma_{\mu\nu}\,S(k)+2T_\mu(k) ~~ . \label{eq:Qsum}
\end{equation}

As previously stated, the term $F_\mu(k,\Delta k)$ given in Eq. (\ref{eq:F1}) is model dependent and cannot be defined in terms of the function $B(k)$, $C(k)$, and $D(k)$. In fact, the function $\rho_\mu(k,\Delta k,q)$ defined in Eq. (\ref{eq:rho1}) depends explicitly on the operator $\hat\Gamma(q)$. However, in order to simplify bookkeeping, we write an expression for $\rho_\mu(k,\Delta k,q)$ using the same coefficients given in Table \ref{tab:coeff}. After considering each interaction separately, we find
\begin{align}
\rho_\mu(k,\Delta k,q) &= -2\Delta k^\nu\left(\left(b_2{k\cdot q\over m_e^2}-d_0\right)q_\nu\gamma_\mu+i\sigma_{\mu\nu}\left(\left(b_2{k\cdot q\over m_e^2}-d_0\right)\slashed q+b_0m_e\right)+{\bar d\over m_e}\slashed qi\sigma_{\mu\nu}\slashed q\right) ~~ . \label{eq:rho2}
\end{align}
Thus, for $F_\mu(k,\Delta k)$ we get
\begin{align}
F_\mu(k,\Delta k) &= {\Delta k^\nu\over2m_e}\left(-\gamma_\mu\left({b_2\over2}I_\nu(k)+{d_0\over2}R\bar I_\nu(k)\right)-i{\bar d\over2}\gamma^\alpha\sigma_{\mu\nu}\gamma^\beta I_{\alpha\beta}(k)\right. \nonumber
\\
& ~~~~~~~~~\left.-i\sigma_{\mu\nu}\left({b_2\over2}\slashed{I}(k)+{d_0\over2}R{\slashed{\bar I}}(k)-{b_0\over2}\left(I_A(k)+{3\over4}R^2\bar I_A(k)\right)\right)\right) ~~ . \label{eq:F2}
\end{align}

\section{phenomenology} \label{pheno}

In this section we translate our results into measured quantities. We take the most up-to-date measurement of the electron $g-2$ found in Ref. \cite{Fan:2022eto} and use it to place constraints on the coupling to the electron of the dark matter candidates considered. The quantity which is actually measured in Refs. \cite{Fan:2022eto,Fan:2022oyb} is $\omega_a=\omega_c-\omega_s$, which ratio $\omega_a/\omega_c$ directly corresponds to the value of $(g-2)/2$. The experimental results of \cite{Fan:2022eto} are in agreement with the standard model prediction to about 
\begin{align}
\delta_E\left(\omega_s-\omega_c\over\omega_c\right)=0.7\times10^{-12}. \label{eq:prec}
\end{align}
For our case, however, the strongest constraints can be determined by comparing our result with the experimental error obtained for $\omega_a$. From Figure (4.27) of \cite{Fan:2022oyb} this error is set to be
\begin{align}
{\delta_E \omega_a\over\omega_a}=4.3\times10^{-12}. \label{eq:prec}
\end{align}

Any dark matter correction for $\omega_a$ that surpasses this value will be constrained\footnote{A more conservative estimate of this error will have a minimal effect on our results since the constraints scale as the square root of the error.}.

To determine $\omega_c$ and $\omega_s$ from our loop correction, we must first find the resulting Hamiltonian for the case of an electron motion in a static and homogeneous magnetic field $\vec B$. Including the correction from the one loop diagrams considered here, such an interaction is described by the following effective lagrangian density 
\begin{align}
{\cal L}_\mathrm{eff}=\bar u(k)\left(\slashed k-m_e+B(k)+\slashed D(k)+A^\mu{\cal M}_\mu^\mathrm{total}\right)u(k) ~~ ,
\end{align}
where $A^\mu$ is the electromagnetic gauge field which has a vanishing zero component $A_0$ and is related to the magnetic field by $\vec A={1\over2}\vec B\times\vec r$. Following our results in Eq. (\ref{eq:Mtot1}) and Eq. (\ref{eq:Qsum}), the electron dynamics is governed by the equation of motion
\begin{align}
\left(\slashed{k}-m_e+B(k)+\slashed{D}(k)-e A^\mu\left(\gamma_\mu+2T_\mu(k)-i{\Delta k^\nu\over m_e}\sigma_{\mu\nu}\,S(k)+i{\Delta k^\rho\over2m_e}{d D^\nu\over dk^\mu}\sigma_{\nu\rho}+F_\mu(k,\Delta k)\right)\right)u(k,s)=0 ~~ ,
\end{align}
where all the terms are given explicitly in Sec. \ref{results}. The Hamiltonian is obtained by solving this equation for $k_0$.


With the definitions $\rho_1=-\gamma_5$, $\rho_2=i\gamma_0\gamma_5$ and $\rho_3=\gamma_0$ \footnote{The definition of $\rho_2$ is necessary for closure of the product of $\rho_i$ operators.}, the Hamiltonian takes the form
\begin{align}
H=-\rho_1\vec\sigma\cdot\vec\pi+\rho_3(m_e-B+S_++S_-)-D_0 ~~ , \label{eq:hamil1}
\end{align}
where $\sigma_i$ are the Pauli matrices and 
\begin{align}
\vec\pi(k)=\vec k-e\vec A\left(1+{D_0(k)\over \omega_k}+b_0m_eR{\bar I_0(k)\over2\omega_k}\right) ~~ . \label{eq:pi1}
\end{align}
The operators $S_+$ and $S_-$ commute and anticommute with $\rho_2$, respectively. They are defined as 
\begin{align}
S_+ &= -{e\over2m_e}\partial^jA^i\sigma_{ij}h_+ ~~ , \label{eq:S+}
\\
S_- &= {e\over2m_e}\partial^jA^i\left(\rho_3\sigma_{ij}h_-^{(1)}-i{k_i\over2m_e}\rho_1\sigma_jh_-^{(2)}\right) ~~ , \label{eq:S-}
\end{align}
where
\begin{align}
h_+ &= -{\bar d\over2}J+{d_0-b_2\over2}RI_A-d_2I_{00}+{1\over3}(d_0/2+d_2)(I_{11}+I_{22}+I_{33}) ~~ , \label{eq:h+}
\\
h_-^{(1)} &= {b_2\over2}I_0+{d_0\over2}R\bar I_0 ~~ , \label{eq:h-1}
\\
h_-^{(2)} &= -\bar dI_0+d_0R\bar I_0 ~~ . \label{eq:h-2}
\end{align}
This result was found under the assumption of a spherically symmetric dark matter distribution, and keeping velocity terms up to the second power. Then, terms of the form $I_i$, $\bar I_i$ and $I_{0i}$ are zero.

\begin{figure}[t]
  \includegraphics[width=0.5\linewidth]{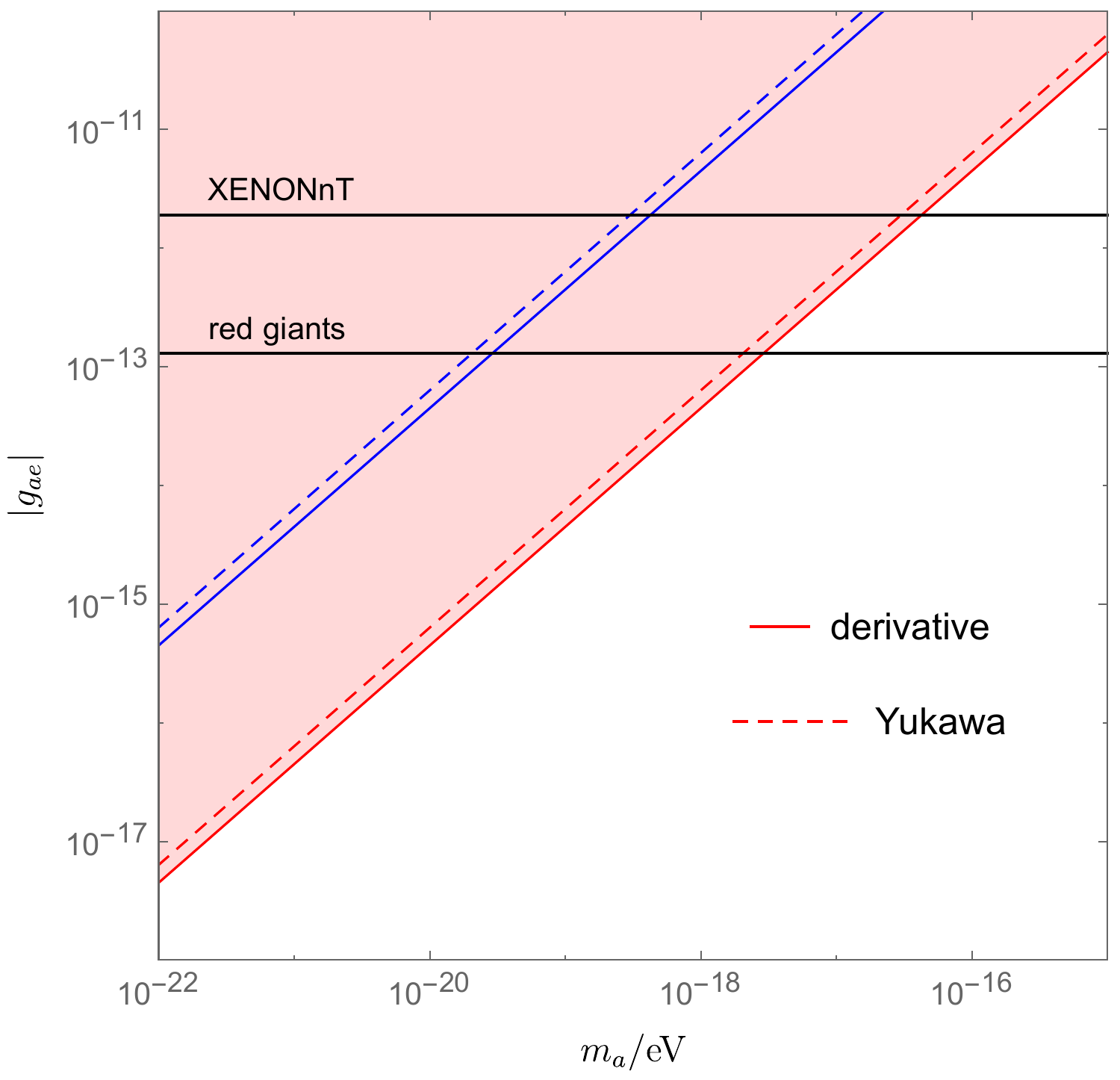}
\caption{Constraints on the axion-electron coupling. The solid red line corresponds to constraints obtained in this work for the case of a derivative coupling while the dashed red line stands for the case of a pseudoscalar Yukawa coupling. The solid and dashed blue lines are constraints for the derivative and Yukawa p-scalar coupling, respectively, assuming the local axion energy density is just a 0.01\% of the total dark matter. The solid black lines are current constraints from solar axions (XENONnT) and red giant evolution.} \label{fig:constraints}
\end{figure}

To diagonalize the Hamiltonian (see section IV in \cite{Donoghue:1984zz}), we perform a Foldy-Wouthuysen transformation\footnote{This procedure only diagonalizes the Dirac equation to leading order. In our final expressions, we have neglected any remaining off diagonal pieces, since they will be suppressed by additional powers of the electron mass.}
\begin{align}
H'=e^{-{i\over2}\phi\rho_2}He^{{i\over2}\phi\rho_2}
\end{align}
such that 
\begin{align}
\tan(\phi)={\vec\sigma\cdot\vec\pi\over m_e-B} ~~ .
\end{align}
After some algebra, we get
\begin{align}
H'=&E_\beta-{e\over2E_\beta}\left(1+b_0m_eR{\bar I_0\over2\omega_k}\right)\vec L\cdot\vec B -{e\over2E_\beta}\left(1+b_0m_eR{\bar I_0\over2\omega_k}+{\omega_k\over m_e}h_+-h_-^{(1)}-{\vec k^2\over4m_e^2}h_-^{(2)}\right)\vec\sigma\cdot\vec B ~~ , \label{eq:hamil2}
\end{align}
where $\vec L$ is the electron angular momentum and $E_\beta=\sqrt{\vec k^2+m_\beta^2}$, with $m_\beta$ the physical mass that the electron acquires from the background. It is given by
\begin{align}
m_\beta=m_e-B-{\omega_k\over m_e}D_0 ~~ .
\end{align}

The cyclotron and spin frequencies are the factors that multiply $\vec L$ and $\vec\sigma$, respectively, in Eq. (\ref{eq:hamil2}), giving
\begin{align}
\omega_c &= {eB\over2E_\beta}\left(1+b_0m_eR{\bar I_0\over2\omega_k}\right) \label{eq:cycfreq}
\\
\omega_s &= \omega_c\left(1+{\alpha\over2\pi}{\omega_k\over m_e}+{\omega_k\over m_e}h_+-h_-^{(1)}-{|\vec k|^2\over4m_e^2}h_-^{(2)}\right) ~~ . \label{eq:spinfreq}
\end{align}
Note, we have added the zero background one-loop quantum electrodynamics contribution.

Now we perform the integrals defined in Eqs. (\ref{eq:J})-(\ref{eq:Imunu}) and input the results into Eqs. (\ref{eq:cycfreq}) and (\ref{eq:spinfreq}). Using the non-relativisitc properties of the dark matter background, we evaluate the integrals after we expand the integrands to second order in the dark matter and electron velocities (See appendix \ref{app:integrals} for details).


Defining
\begin{align}
\lambda={\gae^2\rho_a\over m_a^2m_e^2} ~~ ,
\end{align}
and the dark matter and electron velocities as $v_a$ and $v_e$, respectively, our dark matter background corrections to the parameter $\omega_a$ reduce to the following expressions

\begin{align}
{\delta_T \omega_a\over\omega_a}={2\pi\over\alpha}
\begin{cases}
\lambda\left({1\over3}\left<v_a^2\right>+{1\over8}v_e^2\right) & \text{derivative}
\\
\\
\lambda\left(-{1\over6}\left<v_a^2\right>+{1\over4}v_e^2\right) & \text{pYukawa/sYukawa}
\end{cases} ~~ , \label{eq:predic}
\end{align}
where $\left<\right>$ means averaging over the dark matter velocity distribution, with $f_a(q)$ now a velocity dependent function taken from fits of the dark matter velocity distribution. 


To compare with Eq. (\ref{eq:prec}) we need an idea of the value of $v_e$ in the experiment. The velocity of the electron in the trap are of the order of $v_e\simeq2.3\times 10^{-4} c$ \cite{Fan:2022eto}. As this value is smaller than the dark matter standard average velocity, namely $\sqrt{\left<v_a^2\right>}=270\,\text{km/s}\simeq10^{-3} c$, we neglect the electron velocity contribution.


Making our theoretical predictions in Eq. (\ref{eq:predic}) to be smaller than the experimental error given in Eq. (\ref{eq:prec}), our constraints are

\begin{align}
\gae<
\begin{cases}
4.48\times10^{-14}\left(m_a\over10^{-18}\,\text{eV}\right) & \text{derivative}
\\
\\
6.34\times10^{-14}\left(m_a\over10^{-18}\,\text{eV}\right) & \text{pYukawa/sYukawa}
\end{cases}
\end{align}

In Fig. \ref{fig:constraints}, we compare our constraints for the derivative and pYukawa coupling with the strongest currents constraints. We see that ours are competitive or stronger for masses below $m_a=3\times10^{-18}\,\text{eV}$. For the scalar case, very strong constraints, i.e. $\gae<10^{-24}$ and $10^{-25}$, are given by tests of the equivalency principle \cite{Hees:2018fpg,Berge:2017ovy}. For masses below $m_a=10^{-18}\,\text{eV}$, there are even stronger constraints for scalar dark matter candidates \cite{Kennedy:2020bac,Kobayashi:2022vsf,NANOGrav:2023hvm}. Thus, for the sYukawa case our results are of no effect.

\section{conclusion} \label{conclusion}

In this work we have discussed the effects of a pseudoscalar and scalar dark matter background on the electron $g-2$ value. We have performed loop corrections on the electron-photon vertex using techniques similar to those found in finite temperature field theory. These same techniques were used to perform a similar calculation on the dark photon dark matter case discussed in \cite{Evans:2023uxh}. Taking the most up-to-date measurements of these observables, we were able to place the strongest constraints on axion-like dark matter for axion masses below $3\times10^{-18}\,\text{eV}$. For the case of scalar dark matter, our results are not competitive with current constraints from equivalence principle tests. 

\section*{Acknowledgement}
We would like to thank Tsutomu T. Yanagida and Xing Fan for useful discussions regarding this work. 

\appendix

\section{Renormalization constant} \label{app:renorm}

To calculate the form of the renormalization constant $Z_2$ we first notice that from Eq. (\ref{eq:defktilde}) and Eq. (\ref{eq:defmtilde}) we have
\begin{align}
\tilde k^2-\tilde m_e^2&=((k^0)^2-\vec k^2-m_e^2)\,(1+C)^2+2k^0\,D_0\,(1+C)+D_0^2-2\,\vec k\cdot\vec D\,(1+C)-{(\vec k\cdot\vec D)^2\over\vec k^2}+2\,m_e\,B(1+C)-B^2 \nonumber
\\
&\simeq (1+2C)\,((k^0)^2-\vec k^2-m_e^2+2\,k\cdot D+2\,m_e\,B) \nonumber
\\
&=(1+2C)\,((k^0)^2-E^2), \label{eq:ktminmt}
\end{align}
where we have neglected quadratic background terms, which are higher order, and have defined
\begin{equation}
E(k)=\sqrt{\vec k^2+m_e^2-2\,k\cdot D(k)-2\,m_e\,B(k)}. \label{eq:Eapp}
\end{equation}
We now expand Eq. (\ref{eq:elecprop1}) as
\begin{align}
S_F^R(x-x') &=\int {d^4k\over(2\pi)^4}e^{-ik\cdot(x-x')}{Z_2^{-1}\over\tilde{\slashed{k}}-\tilde m_e+i\epsilon} \nonumber
\\
& = \int{d^3k\over(2\pi)^3}e^{i\vec k\cdot(\vec x-\vec x')}\int{dk_0\over2\pi}{Z_2^{-1}(\tilde{\slashed k}+\tilde m_e)e^{-ik_0(x_0-x_0')}\over\tilde k^2-\tilde m_e^2} \nonumber
\\
&=\int{d^3k\over(2\pi)^3}e^{i\vec k\cdot(\vec x-\vec x')}\int{dk^0\over2\pi}{Z_2^{-1}\over(1+2C)}{(\tilde{\slashed k}+\tilde m_e)e^{-ik^0(x^0-x^{0'})}\over(k^0-\Omega+i\epsilon)(k^0+\Omega-i\epsilon)} \nonumber
\\
&=\int{d^3k\over(2\pi)^3}e^{i\vec k\cdot(\vec x-\vec x')}\left(\Theta(x^0-x^{0'})\int{dk^0\over2\pi}{Z_2^{-1}\over(1+2C)}{\tilde{\slashed k}+\tilde m_e\over2\Omega}e^{-ik^0(x^0-x^{0'})}(-2\pi i)\delta(k^0-\Omega)\right. \nonumber
\\
&\left.~~~+\Theta(x^{0'}-x^0)\int{dk^0\over2\pi}{Z_2^{-1}\over(1+2C)}{\tilde{\slashed k}+\tilde m_e\over-2\Omega}e^{-ik^0(x^0-x^{0'})}2\pi i\,\delta(k^0+\Omega)\right).
\end{align}
The delta functions can be written as
\begin{equation}
\delta(k^0\mp\Omega)={\delta(k^0\mp E)\over1\mp{d\Omega\over dk^0}}={\delta(k^0\mp E)\over1\pm{1\over\Omega}{d\over dk^0}(k\cdot D+m_eB)},
\end{equation}
then
\begin{align}
S_F^R(x-x') &=-i\int{d^3k\over(2\pi)^3}\,e^{i\vec k\cdot\vec x}\left(\Theta(x^0-x^{0'})\left.{Z_2^{-1}\over(1+2C)}{1\over1+{1\over\Omega}{d\over dk^0}(k\cdot D+m_eB)}{\tilde{\slashed k}+\tilde m_e\over2\Omega}\right|_{k^0=E}e^{-iE(x^0-x^{0'})}\right. \nonumber
\\
&\left.~~~~+\Theta(x^{0'}-x^0)\left.{Z_2^{-1}\over(1+2C)}{1\over1-{1\over\Omega}{d\over dk^0}(k\cdot D+m_eB)}{\tilde{\slashed k}+\tilde m_e\over2\Omega}\right|_{k^0=-E}e^{iE(x^0-x^{0'})}\right).
\end{align}
Although not obvious, $C(-k)=C(k)$, $B(-k)=B(k)$, $D(k)=-D(k)$ and $\tilde k(-k)=-\tilde k$ to leading order and we find
\begin{align}
S_F^R(x-x')&=-i\int{d^3k\over(2\pi)^3}{Z_2^{-1}\over(1+2C)}{\tilde E/E\over1+{1\over E}{d\over dE}(k\cdot D+m_eB)}\left(\Theta(x^0-x^{0'}){\tilde{\slashed k}+\tilde m_e\over2E}e^{-ik\cdot x}\right. \nonumber
\\
&\left.~~~~-\Theta(x^{0'}-x^0){\tilde{\slashed k}-\tilde m_e\over2E}e^{ik\cdot x}\right).
\end{align}
Comparing this result with Eq. (\ref{eq:elecprop2}) we get
\begin{align}
Z_2&={1\over(1+2C)}{\tilde E/E\over1+{1\over E}{d\over dE}(k\cdot D+m_eB)} \nonumber
\\
&\simeq(1-2C)\left(1+C+D_0/E\right)\left(1-{1\over E}{d\over dE}(k\cdot D+m_eB)\right) \nonumber
\\
&\simeq1-C-{1\over E}{d\over dE}(k\cdot D+m_eB)+{D_0\over E}.
\end{align}

\section{Computation of the diagrams} \label{app:diagrams}

\subsection{Wave function renormalization}

For the wave function renormalization contribution, we add a factor $Z_2(k)^{-1/2}$ on the spinors $u(k)$ for the zero-order vertex shown on the upper left part of Fig. \ref{fig:diagrams}. We notice that neglecting second order background corrections $Z_2(k)^{-1/2}Z_2(k')^{-1/2}\simeq(Z_2(k)^{-1}+Z_2(k')^{-1})/2$, then
\begin{equation}
i{\cal M}_\mu^{WFR}=(-ie)\left(Z_2(k)^{-1}+Z_2(k')^{-1}\over2\right)\bar u(k')\gamma_\mu u(k). \label{eq:ampWFRapp}
\end{equation}

\subsection{Self-energy contribution}

The self-energy contribution is given by the bottom-right diagram of Fig. \ref{fig:diagrams}. From the Feynman rules we have
\begin{equation}
i{\cal M}_\mu^{SE}=\bar u(k')(-ie\,\gamma_\mu){i\over\slashed{k}-m_e}(i\Sigma_\beta(k))\,u(k),
\end{equation}
where $\Sigma_\beta(k)$ is given by Eq. (\ref{eq:selfen1}). As $\Sigma_\beta(k)$ is not on-shell, we expand $B(k)$ and $D^\mu(k)$ to first order in $k^2-m_e^2$ as
\begin{align}
B(k)&=\left.B(k)\right|_{k^2=m_e^2}+{k^2-m_e^2\over2\omega_k}\left.{\partial B\over\partial E}\right|_{k^2=m_e^2} 
\\
D^\mu(k)&=\left.D^\mu(k)\right|_{k^2=m_e^2}+{k^2-m_e^2\over2\omega_k}\left.{\partial D^\mu\over\partial E}\right|_{k^2=m_e^2},
\end{align}
thus we get
\begin{align}
i{\cal M}_\mu^{SE}&=(-ie)\bar u(k')\,\gamma_\mu\left(-C(k)-{1\over\slashed k-m_e}\left(B(k)+\slashed D(k)\right)-\left(\slashed k+m_e\over2\omega_k\right)\left({\partial B\over\partial E}+{\partial\slashed D\over\partial E}\right)\right)u(k) \nonumber
\\
&=(-ie)\bar u(k')\,\gamma_\mu\left(-C(k)-{1\over\slashed k-m_e}\left(B(k)+\slashed D(k)\right)-{m_e\over\omega_k}{\partial B\over\partial E}-{k\over\omega_k}\cdot{\partial D\over\partial E}\right)u(k) \nonumber
\\
&=(-ie)\bar u(k')\,\gamma_\mu\left(-C(k)-{1\over\slashed k-m_e}\left(B(k)+\slashed D(k)\right)-{m_e\over\omega_k}{\partial B\over\partial E}-{1\over\omega_k}{\partial(k\cdot D)\over\partial E}+{D^0(k)\over\omega_k}\right)u(k) \nonumber
\\
&=(-ie)\bar u(k')\,\gamma_\mu\left(-C(k)-{m_e\over\omega_k}{d\over dE}\left(B+{k\cdot D\over m_e}\right)+{D^0(k)\over\omega_k}-{1\over\slashed k-m_e}\left(B(k)+\slashed D(k)\right)\right)u(k) \nonumber
\\
&=(-ie)\bar u(k')\,\gamma_\mu\left(1-Z_2(k)^{-1}-{1\over\slashed k-m_e}\left(B(k)+\slashed D(k)\right)\right)u(k), \label{eq:ampSEapp}
\end{align}
where $Z_2(k)^{-1}$ is evaluated at $k^2=m_e^2$.

\subsection{Counterterms}

As the electron dynamics is described by the equation $(\slashed k-m_e+\Sigma_\beta(k))u_\beta(k)$, the lagrangian piece that contain the counterterms is
\begin{equation}
{\cal L}^{CT}=-\bar u(k)(B+\slashed D)u(k). \label{eq:lagcount}
\end{equation}
The Feynman rules for the counterterm part gives
\begin{align}
i{\cal M}_\mu^{CT}&=\bar u(k')(-ie\,\gamma_\mu){i\over\slashed{k}-m_e}(-i)(B(k)+\slashed D(k))\,u(k) \nonumber
\\
&=(-ie)\bar u(k')\gamma_\mu{1\over\slashed{k}-m_e}(B(k)+\slashed D(k))\,u(k). \label{eq:ampCTapp}
\end{align}

\subsection{Vertex diagram}

The Feynman amplitude for the vertex diagram is
\begin{equation}
i{\cal M}^{VER}_\mu(k,k') = -ie\bar u(k')\Lambda_\mu(k,k')u(k)~,
\end{equation}
where
\begin{equation}
\Lambda_\mu(k,k')=\gae^2\int{d^4q\over(2\pi)^3}f_a(q)\delta(q^2-m_a^2)i\hat\Gamma(q){i\over\slashed{k'}+\slashed q-m_e}\gamma_\mu{i\over\slashed k+\slashed q-m_e}i\hat\Gamma(q). \label{eq:Lambda}
\end{equation}

Expanding the electron propagator to first order in the photon momentum $\Delta k=k'-k$, we write
\begin{align}
{1\over\slashed{k'}+\slashed{q}-m_e} &= {1\over\slashed k+\slashed{q}+\Delta\slashed k-m_e} \nonumber
\\
&\simeq {1\over\slashed k+\slashed{q}-m_e}+{d\over\Delta k^\mu}\left.\left({1\over\slashed k+\slashed{q}+\Delta\slashed k-m_e}\right)\right|_{\Delta k^\mu=0}\Delta k^\mu \nonumber
\\
&= {1\over\slashed k+\slashed{q}-m_e}-{1\over\slashed k+\slashed{q}-m_e}\Delta\slashed k{1\over\slashed k+\slashed{q}-m_e} ~~ , \label{eq:propexpk'}
\\
{1\over\slashed k+\slashed{q}-m_e} &= {1\over\slashed{k'}+\slashed{q}-\Delta\slashed k-m_e} \nonumber
\\
&\simeq {1\over\slashed{k'}+\slashed{q}-m_e}+{d\over\Delta k^\mu}\left.\left({1\over\slashed{k'}+\slashed{q}-\Delta\slashed k-m_e}\right)\right|_{\Delta k^\mu=0}\Delta k^\mu \nonumber
\\
&= {1\over\slashed{k'}+\slashed{q}-m_e}+{1\over\slashed{k'}+\slashed{q}-m_e}\Delta\slashed k{1\over\slashed{k'}+\slashed{q}-m_e}
\\
&\simeq {1\over\slashed{k'}+\slashed{q}-m_e}+{1\over\slashed{k}+\slashed{q}-m_e}\Delta\slashed k{1\over\slashed{k}+\slashed{q}-m_e}~~ . \label{eq:propexpk}
\end{align}
Thus, replacing Eq. (\ref{eq:propexpk'}) into Eq. (\ref{eq:Lambda}) and then Eq. (\ref{eq:propexpk}) into Eq. (\ref{eq:Lambda}) we have
\begin{equation}
\Lambda_\mu(k,k')=\Lambda_\mu(k,k)-\gae^2\int{d^4q\over(2\pi)^3}f_a(q)\delta(q^2-m_a^2)i\hat\Gamma(q){i\over\slashed{k}+\slashed q-m_e}\Delta\slashed{k}{1\over\slashed k+\slashed q-m_e}\gamma_\mu{i\over\slashed{k}+\slashed q-m_e}i\hat\Gamma(q) ~~ \label{eq:Lambda1}
\end{equation}
and
\begin{equation}
\Lambda_\mu(k,k')=\Lambda_\mu(k',k')+\gae^2\int{d^4q\over(2\pi)^3}f_a(q)\delta(q^2-m_a^2)i\hat\Gamma(q){i\over\slashed k+\slashed q-m_e}\gamma_\mu{1\over\slashed{k}+\slashed q-m_e}\Delta\slashed{k}{i\over\slashed{k}+\slashed q-m_e}i\hat\Gamma(q) ~~ , \label{eq:Lambda2}
\end{equation}
respectively. Now, combining both Eq. (\ref{eq:Lambda1}) and Eq. (\ref{eq:Lambda2}) we get
\begin{equation}
\Lambda_\mu(k,k')={1\over2}\left(\Lambda_\mu(k,k)+\Lambda_\mu(k',k')\right)+{\gae^2\over2}\int{d^4q\over(2\pi)^3}f_a(q)\delta(q^2-m_a^2)i\hat\Gamma(q)\eta_\mu(k,\Delta k,q)i\hat\Gamma(q) ~~ ,
\end{equation}
where
\begin{equation}
\eta_\mu(k,\Delta k,q)={i\over\slashed k+\slashed q-m_e}\left(\gamma_\mu{1\over\slashed k+\slashed q-m_e}\Delta\slashed k-\Delta\slashed k{1\over\slashed k+\slashed q-m_e}\gamma_\mu\right){i\over\slashed k+\slashed q-m_e} ~~ .
\end{equation}
This function can be reduced as
\begin{align}
\eta_\mu(k,\Delta k,q) &= {1\over(k+q)^2-m_e^2}{i\over\slashed k+\slashed q-m_e}\left(\gamma_\mu\,(\slashed k+\slashed q+m_e)\,\Delta\slashed k-\Delta\slashed k\,(\slashed k+\slashed q+m_e)\,\gamma_\mu\right){i\over\slashed k+\slashed q-m_e} \nonumber
\\
&= {1\over(k+q)^2-m_e^2}{i\over\slashed k+\slashed q-m_e}\left(2(k+q)\cdot\Delta k\,\gamma_\mu-\gamma_\mu\Delta\slashed k\,(\slashed k+\slashed q-m_e)\right. \nonumber
\\
& ~~~~~~~~~~~~~~~~~~~~~~~~~~~~~~~~~~~~~\left.-2\Delta k\cdot(k+q)\gamma_\mu+(\slashed k+\slashed q-m_e)\Delta\slashed k\,\gamma_\mu\right){i\over\slashed k+\slashed q-m_e} \nonumber
\\
&= {1\over(k+q)^2-m_e^2}\left({1\over\slashed k+\slashed q-m_e}\gamma_\mu\Delta\slashed k-\Delta\slashed k\,\gamma_\mu{1\over\slashed k+\slashed q-m_e}\right) \nonumber
\\
&= {1\over((k+q)^2-m_e^2)^2}\left((\slashed k+\slashed q+m_e)\gamma_\mu\Delta\slashed k-\Delta\slashed k\,\gamma_\mu(\slashed k+\slashed q+m_e)\right) \nonumber
\\
&= {1\over((k+q)^2-m_e^2)^2}\left((\slashed k+\slashed q)\gamma_\mu\Delta\slashed k-\Delta\slashed k\,\gamma_\mu(\slashed k+\slashed q)+m_e[\gamma_\mu,\Delta\slashed k]\right) ~~.
\end{align}
We now define the function $\rho_\mu(k,\Delta k,q)$ such that
\begin{equation}
i\hat\Gamma(q)\eta_\mu(k,\Delta k,q)i\hat\Gamma(q)={\rho_\mu(k,\Delta k,q)\over((k+q)^2-m_e^2)^2} ~~ ,
\end{equation}
then we express it as
\begin{align}
\rho_\mu(k,\Delta k,q) &= i\hat\Gamma(q)\left((\slashed k+\slashed q)\gamma_\mu\Delta\slashed k-\Delta\slashed k\,\gamma_\mu(\slashed k+\slashed q)+m_e[\gamma_\mu,\Delta\slashed k]\right)i\hat\Gamma(q) \nonumber
\\
&= [i\hat\Gamma(q),\slashed k]\,\gamma_\mu\,\Delta\slashed k\,i\hat\Gamma(q)+i\hat\Gamma(q)\,\Delta\slashed k\,\gamma_\mu\,[i\hat\Gamma(q),\slashed k] \nonumber
\\
& ~~~ +i\hat\Gamma(q)(\slashed q\,\gamma_\mu\,\Delta\slashed k-\Delta\slashed k\,\gamma_\mu\,\slashed q)i\hat\Gamma(q)+2m_ei\hat\Gamma(q)[\gamma_\mu,\Delta\slashed k]i\hat\Gamma(q) ~~ ,
\end{align}
where to get the second contribution with the $[\gamma_\mu,\Delta\slashed k]$ matrix, we have used $(\slashed k-m_e)u(k)=0$ and $\bar u(k)(\slashed k-m_e)=0$. Thus, the vertex diagram amplitude can be expressed as
\begin{equation}
\Lambda_\mu(k,k')={1\over2}\left(\Lambda_\mu(k,k)+\Lambda_\mu(k',k')\right)+F_\mu(k,\Delta k) ~~ ,
\end{equation}
where
\begin{equation}
F_\mu(k,\Delta k)={\gae^2\over2}\int{d^4q\over(2\pi)^3}f_a(q)\delta(q^2-m_a^2){\rho_\mu(k,\Delta k,q)\over((k+q)^2-m_e^2)^2} ~~ .
\end{equation}

Finally, by using the identity ${d\over dx}A^{-1}=-A^{-1}{dA\over dx}A^{-1}$, we notice from Eq. (\ref{eq:selfen0}) that
\begin{align}
\Lambda_\mu(k,k) &= \gae^2\int{d^4q\over(2\pi)^3}f_a(q)\delta(q^2-m_a^2)i\hat\Gamma(q){i\over\slashed k+\slashed q-m_e}\gamma_\mu{i\over\slashed{k}+\slashed q-m_e}i\hat\Gamma(q)
\\
&= \gae^2\int{d^4q\over(2\pi)^3}f_a(q)\delta(q^2-m_a^2)i\hat\Gamma(q){d\over dk^\mu}\left({1\over\slashed k+\slashed q-m_e}\right)i\hat\Gamma(q)
\\
&= {d\Sigma_\beta\over dk^\mu} ~~ .
\end{align}
In terms of the general functions $B(k)$, $C(k)$ and $D_\mu(k)$, we have
\begin{equation}
{d\Sigma_\beta\over dk^\mu}={dB\over dk^\mu}+{dC\over dk^\mu}(\slashed k-m_e)+C(k)\gamma_\mu+\gamma^\nu{dD_\nu\over dk^\mu} ~~ .
\end{equation}
Now, using the Gordon identity $\gamma^\nu={k^\nu+k'^\nu\over2m_e}+{i\Delta k_\rho\over2m_e}\sigma^{\nu\rho}$, where $\sigma^{\mu\nu}={i\over2}[\gamma^\mu,\gamma^\nu]$, we write
\begin{align}
\gamma^\nu{dD_\nu\over dk^\mu} &= {k^\nu\over m_e}{dD_\nu\over dk^\mu}+{\Delta k^\nu\over2m_e}{dD_\nu\over dk^\mu}+{i\Delta k_\rho\over2m_e}\sigma^{\nu\rho}{dD_\nu\over dk^\mu} \nonumber
\\
&= {d\over dk^\mu}\left(k\cdot D\over m_e\right)-{D_\mu(k)\over m_e}+{\Delta k^\nu\over2m_e}{dD_\nu\over dk^\mu}+{i\Delta k_\rho\over2m_e}\sigma^{\nu\rho}{dD_\nu\over dk^\mu} ~~ ,
\end{align}
and then
\begin{equation}
{d\Sigma_\beta\over dk^\mu}=C(k)\gamma_\mu+{d\over dk^\mu}\left(B+{k\cdot D\over m_e}\right)-{D_\mu(k)\over m_e}+{i\Delta k_\rho\over2m_e}\sigma^{\nu\rho}{dD_\nu\over dk^\mu}+{\Delta k^\nu\over2m_e}{dD_\nu\over dk^\mu} ~~ ,
\end{equation}
where the ${dC\over dk^\mu}$ term has been removed by using $(\slashed k-m_e)u(k)=0$. Analogously, we find
\begin{equation}
{d\Sigma_\beta\over dk'^\mu}=C(k')\gamma_\mu+{d\over dk'^\mu}\left(B+{k'\cdot D\over m_e}\right)-{D_\mu(k')\over m_e}+{i\Delta k_\rho\over2m_e}\sigma^{\nu\rho}{dD_\nu\over dk'^\mu}-{\Delta k^\nu\over2m_e}{dD_\nu\over dk'^\mu} ~~ .
\end{equation}

Our final general expression for the vertex diagram is
\begin{align}
\Lambda_\mu(k,k') &= {1\over2}(C(k)+C(k'))\gamma_\mu+{1\over 2}{d\over dk^\mu}\left(B+{k\cdot D\over m_e}\right)+{1\over2}{d\over dk'^\mu}\left(B+{k'\cdot D\over m_e}\right) \nonumber
\\
& ~~~ -{D_\mu(k)+D_\mu(k')\over2m_e}+{i\Delta k_\rho\over2m_e}\sigma^{\nu\rho}{dD_\nu\over dk^\mu}+F_\mu(k,\Delta k) ~~ .
\end{align}

\section{Computation of the integrals} \label{app:integrals}

In this section we deal with the integrals defined between Eq. (\ref{eq:J}) and Eq. (\ref{eq:Imunu}). For this, we assume a spherically symmetric distribution $f_a(q)$ and because of the non relativistic properties of the axion and electron motion, we also neglect third power velocity terms and higher orders.

The integrals have a general form given by
\begin{align}
I_{\alpha,n}(k) &= \int {d^4q\over(2\pi)^3}f_a(q)\delta(q^2-m_a^2){\alpha(q)\over(q\cdot k)^n} ~~ . \label{eq:int1}
\end{align}
Notice that the combination $\alpha(q)\over(q\cdot k)^n$ is also symmetric under $q\rightarrow-q$, at least for all the functions considered in our calculation. By using the fact that $\delta(q^2-m_a^2)={1\over\Omega_q}(\delta(q_0-\Omega_q)+\delta(q_0+\Omega_q))$, where $\Omega_q=\sqrt{\vec q^2+m_a^2}$, we have
\begin{align}
I_{\alpha,n}(k) &= \gae^2\int {d^3q\over(2\pi)^3}{1\over\Omega_q}\left({f_a(\Omega_q,\vec q)\,\alpha(\Omega_q,\vec q)\over\left(\Omega_q\,\omega_k-\vec q\cdot\vec k\right)^n}+{f_a(-\Omega_q,\vec q)\,\alpha(-\Omega_q,\vec q)\over\left(-\Omega_q\,\omega_k-\vec q\cdot\vec k\right)^n}\right) \nonumber
\\
&= \gae^2\int {d^3q\over(2\pi)^3}{f_a(\Omega_q,\vec q)\,\alpha(\Omega_q,\vec q)\over\Omega_q\left(\Omega_q\,\omega_k-\vec q\cdot\vec k\right)^n} ~~ , \label{eq:int2}
\end{align}
where in the last step we made $\vec q\rightarrow-\vec q$ in the second term and used the symmetry properties of the whole integrand. As $f_a(q)$ is a probability distribution over the phase space, the integral is simply reduced to
\begin{align}
I_{\alpha,n}(k) &= \gae^2{\rho_a\over m_a}\left<{\alpha(\vec q)\over\Omega_q\left(\Omega_q\,\omega_k-\vec q\cdot\vec k\right)^n}\right> ~~ , \label{eq:int3}
\end{align}
where $\rho_a$ is the local axion energy density and $\left<\right>$ means averaging over the dark matter velocity (or momentum distribution).

Next, we expand the denominator of the averaged term in powers of the axion and electron velocities $\vec v_a$ and $\vec v_e$. We have
\begin{align}
{1\over\Omega_q\left(\Omega_q\,\omega_k-\vec q\cdot\vec k\right)^n} &= {1\over\Omega_q^{n+1}\,\omega_k^n}{1\over\left(1-\vec v_a\cdot\vec v_e\right)^n}  \nonumber
\\
&\simeq {1\over m_a^{n+1}\,m_e^n}\left(1-{n+1\over2}v_a^2\right)\left(1-{n\over2}v_e^2\right)\left(1+n\,\vec v_a\cdot\vec v_e\right) \nonumber
\\
&\simeq {1\over m_a^{n+1}\,m_e^n}\left(1-{n+1\over2}v_a^2-{n\over2}v_e^2+n\,\vec v_a\cdot\vec v_e\right) ~~ . \label{eq:velexp}
\end{align}
Then, the final expression for the integrals become
\begin{align}
I_{\alpha,n}(k) &= \lambda {m_e^{2-n}\over m_a^n}\left<\left(1-{n+1\over2}v_a^2-{n\over2}v_e^2+n\,\vec v_a\cdot\vec v_e\right)\alpha(\vec q)\right> ~~ , \label{eq:int4}
\end{align}
where $\lambda=\gae^2{\rho_a\over m_a^2m_e^2}$.

Now we calculate each integral explicitly, considering that $\left<\vec v_a\right>=0$ under the assumption of a spherically symmetric distribution. For $J$ we have $n=0$ and $\alpha(\vec q)=1/m_e^2$, then
\begin{equation}
J = \lambda\left(1-{1\over2}\left<v_a^2\right>\right) ~~ . \label{eq:Jres}
\end{equation}
For $I_A$, $n=2$ and $\alpha(\vec q)=m_e^2$, then
\begin{equation}
I_A = {\lambda\over R}\left(1-{3\over2}\left<v_a^2\right>-v_e^2\right) ~~ . \label{eq:IAres}
\end{equation}
For $\bar I_A$, $n=4$ and $\alpha(\vec q)=m_e^6$, so we get
\begin{equation}
\bar I_A = {\lambda\over R^2}\left(1-{5\over2}\left<v_a^2\right>-2v_e^2\right) ~~ . \label{eq:IAbarres}
\end{equation}
For $I_\mu$, $n=1$ and $\alpha_\mu(\vec q)=q_\mu/m_e$. We have
\begin{equation}
I_0 = \lambda\left(1-{1\over2}\left<v_a^2\right>-{1\over2}v_e^2\right)  \label{eq:I0res}
\end{equation}
and
\begin{equation}
I_i = \lambda\left<v_{ai}\right> \simeq 0 ~~ . \label{eq:Iires}
\end{equation}
For $\bar I_\mu$, $n=3$ and $\alpha_\mu(\vec q)=m_e^3\,q_\mu$. We find
\begin{equation}
\bar I_0 = {\lambda\over R}\left(1-{3\over2}\left<v_a^2\right>-{3\over2}v_e^2\right) \label{eq:I0barres}
\end{equation}
and
\begin{equation}
\bar I_i = {\lambda\over R}\left<v_{ai}\right> \simeq 0 ~~ . \label{eq:Iibarres}
\end{equation}
Finally, for $I_{\mu\nu}$, $n=2$ and $\alpha_{\mu\nu}(\vec q)=q_\mu q_\nu$. We obtain
\begin{equation}
I_{00} = \lambda\left(1-{1\over2}\left<v_a^2\right>-v_e^2\right) ~~ ,  \label{eq:I00res}
\end{equation}
\begin{equation}
I_{0i} = \lambda\left<v_{ai}\right> \simeq 0  \label{eq:I0ires}
\end{equation}
and
\begin{align}
I_i^j = \lambda\left<v_{ai}v_{aj}\right> = -\delta_i^j\lambda\left<v_{ai}^2\right> = -{\delta_i^j\over 3}\lambda\left<v_{a}^2\right> ~~ . \label{eq:Iiires}
\end{align}

\bibliographystyle{JHEP}
\bibliography{references.bib}

\providecommand{\href}[2]{#2}\begingroup\justify\begin{thebibliography}{10}

\bibitem{Peccei:1977hh}
R.~D. Peccei and H.~R. Quinn, \emph{{CP Conservation in the Presence of
  Instantons}},
  \href{https://dx.doi.org/10.1103/PhysRevLett.38.1440}{\emph{Phys. Rev. Lett.}
  {\bf 38} (1977) 1440--1443}.

\bibitem{Kim:1979if}
J.~E. Kim, \emph{{Weak Interaction Singlet and Strong CP Invariance}},
  \href{https://dx.doi.org/10.1103/PhysRevLett.43.103}{\emph{Phys. Rev. Lett.}
  {\bf 43} (1979) 103}.

\bibitem{Shifman:1979if}
M.~A. Shifman, A.~I. Vainshtein and V.~I. Zakharov, \emph{{Can Confinement
  Ensure Natural CP Invariance of Strong Interactions?}},
  \href{https://dx.doi.org/10.1016/0550-3213(80)90209-6}{\emph{Nucl. Phys. B}
  {\bf 166} (1980) 493--506}.

\bibitem{Dine:1981rt}
M.~Dine, W.~Fischler and M.~Srednicki, \emph{{A Simple Solution to the Strong
  CP Problem with a Harmless Axion}},
  \href{https://dx.doi.org/10.1016/0370-2693(81)90590-6}{\emph{Phys. Lett. B}
  {\bf 104} (1981) 199--202}.

\bibitem{Zhitnitsky:1980tq}
A.~R. Zhitnitsky, \emph{{On Possible Suppression of the Axion Hadron
  Interactions. (In Russian)}}, {\emph{Sov. J. Nucl. Phys.} {\bf 31} (1980)
  260}.

\bibitem{Svrcek:2006yi}
P.~Svrcek and E.~Witten, \emph{{Axions In String Theory}},
  \href{https://dx.doi.org/10.1088/1126-6708/2006/06/051}{\emph{JHEP} {\bf 06}
  (2006) 051} [\href{https://arxiv.org/abs/hep-th/0605206}{{\tt
  arXiv:hep-th/0605206}}].

\bibitem{Arvanitaki:2009fg}
A.~Arvanitaki, S.~Dimopoulos, S.~Dubovsky, N.~Kaloper and J.~March-Russell,
  \emph{{String Axiverse}},
  \href{https://dx.doi.org/10.1103/PhysRevD.81.123530}{\emph{Phys. Rev. D} {\bf
  81} (2010) 123530} [\href{https://arxiv.org/abs/0905.4720}{{\tt
  arXiv:0905.4720}}].

\bibitem{Preskill:1982cy}
J.~Preskill, M.~B. Wise and F.~Wilczek, \emph{{Cosmology of the Invisible
  Axion}}, \href{https://dx.doi.org/10.1016/0370-2693(83)90637-8}{\emph{Phys.
  Lett. B} {\bf 120} (1983) 127--132}.

\bibitem{Abbott:1982af}
L.~Abbott and P.~Sikivie, \emph{{A Cosmological Bound on the Invisible Axion}},
  \href{https://dx.doi.org/10.1016/0370-2693(83)90638-X}{\emph{Phys. Lett. B}
  {\bf 120} (1983) 133--136}.

\bibitem{Dine:1982ah}
M.~Dine and W.~Fischler, \emph{{The Not So Harmless Axion}},
  \href{https://dx.doi.org/10.1016/0370-2693(83)90639-1}{\emph{Phys. Lett. B}
  {\bf 120} (1983) 137--141}.

\bibitem{Arias:2012az}
P.~Arias, D.~Cadamuro, M.~Goodsell, J.~Jaeckel, J.~Redondo and A.~Ringwald,
  \emph{{WISPy Cold Dark Matter}},
  \href{https://dx.doi.org/10.1088/1475-7516/2012/06/013}{\emph{JCAP} {\bf 06}
  (2012) 013} [\href{https://arxiv.org/abs/1201.5902}{{\tt arXiv:1201.5902}}].

\bibitem{Sikivie:1983ip}
P.~Sikivie, \emph{{Experimental Tests of the Invisible Axion}},
  \href{https://dx.doi.org/10.1103/PhysRevLett.51.1415}{\emph{Phys. Rev. Lett.}
  {\bf 51} (1983) 1415--1417}.

\bibitem{Sikivie:2020zpn}
P.~Sikivie, \emph{{Invisible Axion Search Methods}},
  \href{https://dx.doi.org/10.1103/RevModPhys.93.015004}{\emph{Rev. Mod. Phys.}
  {\bf 93} (2021) 015004} [\href{https://arxiv.org/abs/2003.02206}{{\tt
  arXiv:2003.02206}}].

\bibitem{Irastorza:2018dyq}
I.~G. Irastorza and J.~Redondo, \emph{{New experimental approaches in the
  search for axion-like particles}},
  \href{https://dx.doi.org/10.1016/j.ppnp.2018.05.003}{\emph{Prog. Part. Nucl.
  Phys.} {\bf 102} (2018) 89--159}
  [\href{https://arxiv.org/abs/1801.08127}{{\tt arXiv:1801.08127}}].

\bibitem{XENON:2019gfn}
{\scshape XENON}, E.~Aprile et~al., \emph{{Light Dark Matter Search with
  Ionization Signals in XENON1T}},
  \href{https://dx.doi.org/10.1103/PhysRevLett.123.251801}{\emph{Phys. Rev.
  Lett.} {\bf 123} (2019) 251801} [\href{https://arxiv.org/abs/1907.11485}{{\tt
  arXiv:1907.11485}}].

\bibitem{XENON:2022ltv}
{\scshape XENON}, E.~Aprile et~al., \emph{{Search for New Physics in Electronic
  Recoil Data from XENONnT}},
  \href{https://dx.doi.org/10.1103/PhysRevLett.129.161805}{\emph{Phys. Rev.
  Lett.} {\bf 129} (2022) 161805} [\href{https://arxiv.org/abs/2207.11330}{{\tt
  arXiv:2207.11330}}].

\bibitem{Ferreira:2022egk}
R.~Z. Ferreira, M.~C.~D. Marsh and E.~M\"uller, \emph{{Do Direct Detection
  Experiments Constrain Axionlike Particles Coupled to Electrons?}},
  \href{https://dx.doi.org/10.1103/PhysRevLett.128.221302}{\emph{Phys. Rev.
  Lett.} {\bf 128} (2022) 221302} [\href{https://arxiv.org/abs/2202.08858}{{\tt
  arXiv:2202.08858}}].

\bibitem{Capozzi:2020cbu}
F.~Capozzi and G.~Raffelt, \emph{{Axion and neutrino bounds improved with new
  calibrations of the tip of the red-giant branch using geometric distance
  determinations}},
  \href{https://dx.doi.org/10.1103/PhysRevD.102.083007}{\emph{Phys. Rev. D}
  {\bf 102} (2020) 083007} [\href{https://arxiv.org/abs/2007.03694}{{\tt
  arXiv:2007.03694}}].

\bibitem{Evans:2023uxh}
J.~L. Evans, \emph{{Effect of Ultralight Dark Matter on $g-2$ of the
  Electron}},  \href{https://arxiv.org/abs/2302.08746}{{\tt arXiv:2302.08746}}.

\bibitem{Donoghue:1984zz}
J.~F. Donoghue, B.~R. Holstein and R.~W. Robinett, \emph{{QUANTUM
  ELECTRODYNAMICS AT FINITE TEMPERATURE}},
  \href{https://dx.doi.org/10.1016/0003-4916(85)90016-8}{\emph{Annals Phys.}
  {\bf 164} (1985) 233}.

\bibitem{Donoghue:1983qx}
J.~F. Donoghue and B.~R. Holstein, \emph{{Renormalization and Radiative
  Corrections at Finite Temperature}},
  \href{https://dx.doi.org/10.1103/PhysRevD.29.3004}{\emph{Phys. Rev. D} {\bf
  28} (1983) 340}.

\bibitem{Fan:2022eto}
X.~Fan, T.~G. Myers, B.~A.~D. Sukra and G.~Gabrielse, \emph{{Measurement of the
  Electron Magnetic Moment}},
  \href{https://dx.doi.org/10.1103/PhysRevLett.130.071801}{\emph{Phys. Rev.
  Lett.} {\bf 130} (2023) 071801} [\href{https://arxiv.org/abs/2209.13084}{{\tt
  arXiv:2209.13084}}].

\bibitem{Fan:2022oyb}
X.~Fan, \emph{{An Improved Measurement of the Electron Magnetic Moment}}.
\newblock PhD thesis, Harvard U., 2022.

\bibitem{Hees:2018fpg}
A.~Hees, O.~Minazzoli, E.~Savalle, Y.~V. Stadnik and P.~Wolf, \emph{{Violation
  of the equivalence principle from light scalar dark matter}},
  \href{https://dx.doi.org/10.1103/PhysRevD.98.064051}{\emph{Phys. Rev. D} {\bf
  98} (2018) 064051} [\href{https://arxiv.org/abs/1807.04512}{{\tt
  arXiv:1807.04512}}].

\bibitem{Berge:2017ovy}
J.~Berg\'e, P.~Brax, G.~M\'etris, M.~Pernot-Borr\`as, P.~Touboul and J.-P.
  Uzan, \emph{{MICROSCOPE Mission: First Constraints on the Violation of the
  Weak Equivalence Principle by a Light Scalar Dilaton}},
  \href{https://dx.doi.org/10.1103/PhysRevLett.120.141101}{\emph{Phys. Rev.
  Lett.} {\bf 120} (2018) 141101} [\href{https://arxiv.org/abs/1712.00483}{{\tt
  arXiv:1712.00483}}].

\bibitem{Kennedy:2020bac}
C.~J. Kennedy, E.~Oelker, J.~M. Robinson, T.~Bothwell, D.~Kedar, W.~R. Milner
  et~al., \emph{{Precision Metrology Meets Cosmology: Improved Constraints on
  Ultralight Dark Matter from Atom-Cavity Frequency Comparisons}},
  \href{https://dx.doi.org/10.1103/PhysRevLett.125.201302}{\emph{Phys. Rev.
  Lett.} {\bf 125} (2020) 201302} [\href{https://arxiv.org/abs/2008.08773}{{\tt
  arXiv:2008.08773}}].

\bibitem{Kobayashi:2022vsf}
T.~Kobayashi et~al., \emph{{Search for Ultralight Dark Matter from Long-Term
  Frequency Comparisons of Optical and Microwave Atomic Clocks}},
  \href{https://dx.doi.org/10.1103/PhysRevLett.129.241301}{\emph{Phys. Rev.
  Lett.} {\bf 129} (2022) 241301} [\href{https://arxiv.org/abs/2212.05721}{{\tt
  arXiv:2212.05721}}].

\bibitem{NANOGrav:2023hvm}
{\scshape NANOGrav}, A.~Afzal et~al., \emph{{The NANOGrav 15 yr Data Set:
  Search for Signals from New Physics}},
  \href{https://dx.doi.org/10.3847/2041-8213/acdc91}{\emph{Astrophys. J. Lett.}
  {\bf 951} (2023) L11} [\href{https://arxiv.org/abs/2306.16219}{{\tt
  arXiv:2306.16219}}].

\end{thebibliography}\endgroup

\end{document}